# A Chatbot for Information Security

**Sofian Hamad, Taoufik Yeferny**
*sofian.hamad@nbu.edu.sa*
Computer Science Department, Northern Border University  Arar Kingdom of Saudi Arabia

**Abstract**
Advancements in artificial intelligence (AI), speech recognition systems (ASR), and machine learning have enabled the development of intelligent computer programs called chatbots. Many chatbots have been proposed to provide different services in many areas such as customer service, sales and marketing. However, the use of chatbot as advisers in the field of information security is not yet considered. Furthermore, people, especially normal users who have no technical background, are unaware about many of aspects in information security. Therefore, in this paper we proposed a chatbot that acts as an adviser in information security. The proposed adviser uses a knowledge base with json file. Having such chatbot provides many features including raising the awareness in field of information security by offering accurate advice, based on different opinions from information security expertise, for many users on different. Furthermore, this chatbot is currently deployed through Telegram platform, which is one of widely used social network platforms. The deployment of the proposed chatbot over different platforms is considered as the future work.

## 1. Introduction

Recent advancements in the technology have supported the emerging and proliferation of ChatBot, which is an intelligent computer program that chats with people. Chatbots can play a role of virtual adviser that uses concepts of automatic speech recognition systems (ASR), machine learning and artificial intelligence (AI). Natural Language Processing (NLP) provides a simple interface between users and a Chatbot as shown in figure 1 [1].

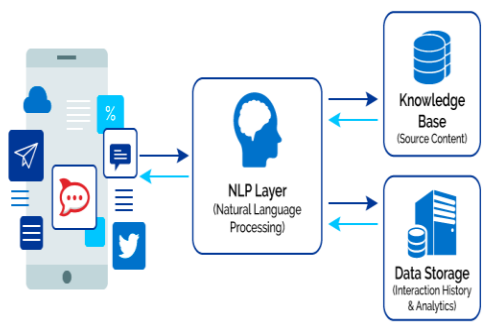

Fig. 1  Chatbot developed by NLP

Furthermore, the use of technologies such as NLP, makes virtual advisers in any field to efficiently and effectively (similar to a real adviser) speak with users. From the other hand, reference [2] has presented a brief introduction on cloud-based chatbots technologies along with programming of chatbots and challenges of programming in current and future Era of chatbot. Examples of cloud based Chatbots services include the chatbot sector such as IBM Watson, Microsoft bot, AWS Lambda, Heroku and many others.

Chatbots (which is also called "digital assistants") can chat with users in many different ways including text-based entities, voice user interfaces, and embodied conversational entities [3]. At the beginning, dialog agents intended to simulate human conversation.

To raise the awareness of normal users, who increasingly use technologies, who have no background about the necessary information security aspects to protect themselves, this paper introduces a chatbot application (text-based entities) to provide information security advice for users. In this paper, we will develop a back end interface that interact with a knowledge base. This knowledge base represents knowledge of expertise in the field of information security. The chatbot is able to understand users' questions and enquires. It is important to note that in this paper, the proposed chatbot extracts keywords and finds matching replies from the knowledge base. In addition, the proposed chatbotis expected to be deployed over one platform, which is Telegram.

In this paper, section 1 presented general   introduction, while section 2 provide the related research. The description of the proposed solution is presented in section 3, while the conclusion is presented in the last section, section 4.

## 2. Background and related work

A. Overview

A chatbot, which is known as interactive agent, conversational interface, Conversational AI, or artificial conversational entity, is a computer program using technology of an artificial intelligence, that makes a conversation through different ways such as text-based ways. These computer programs behave similar to humans when then involved in a conversation.

Chatbots can be developed to extract some keywords during a dialog with users, search in its database and finally send their replies. From the other hand, some Chatbots employ natural language processing systems, which is complex technology, to interact with users.

Leading companies have also used Chatbots, e.g., Google Assistant and Amazon Alexa, and though many platforms including Facebook Messenger, or applications and websites of individual organizations.

The ability of a computer program to act as a human during live written conversation is examined early since 1950 [5]. The results (based on the content of conversation only) show that a computer program and a human responses are similar in their conversational behavior, cannot be distinguished reliably.

B. Maintaining the Integrity of the Specifications

One of the early intelligent computer programs is called ELIZA, which is developed in 1960s and has used natural language processing. ELIZA has showed that communication between humans and machines involving only the most obvious things.

Another chatbot inspired by ELIZA program is the artificial linguistic internet computer entity (A.L.I.C.E.), which is natural language processing chatterbot.

Providers of Platform as a Service (PaaS) such as Oracle Cloud Platform offers many tools to develop, to test and deploy intelligent programs (chatbots). This can performed through cloud[6]. Cloud platforms support many technologies such artificial intelligence, natural language processing, and mobile backend and offer them as a Service.

C. Similar Chatbot

In this section, we have discussed the solutions (chatbots) that have been developed in many different fields. In each of these solutions, additional features (services) is needed to consider the nature of each field.

In [7], authors have developed two working chatbots using two different programming languages (C++ and AIML), to study their construction and design practices thoroughly and to devise the further possible improvements in such kind of programs. They (authors)have discussed Artificial Intelligence via chat bots, their framework design, capabilities, utilization and future scope.

Chatbot is utilized to provide counseling service in many fields such as education [8], insurance services [9], University admission services[10], ontology [11],

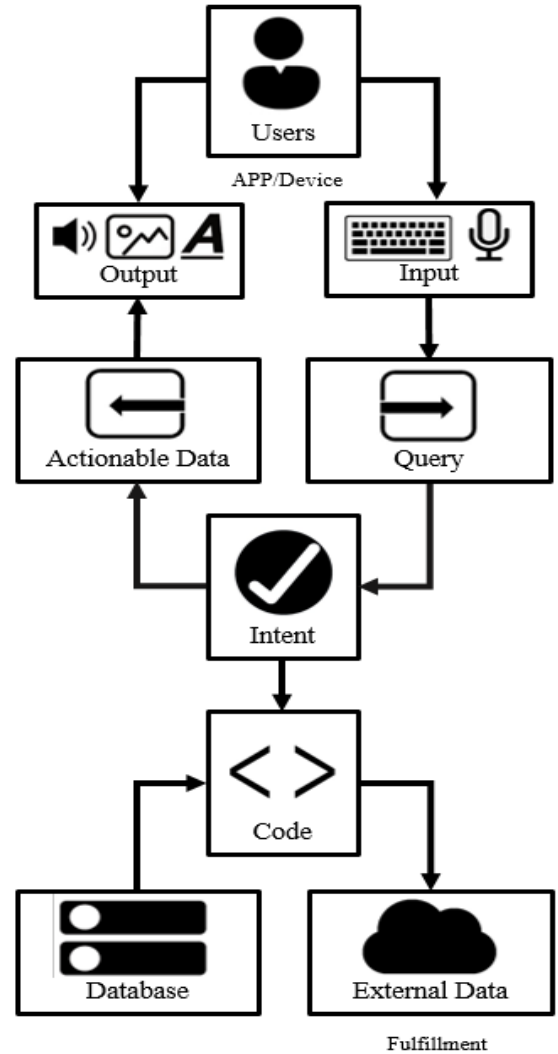

Fig. 2 MedBot Artitechiture

Medical Consultant System [12], psychiatric counseling service[13] where authors use additional constrains to generation model for the proper response generation, which can detect conversational context, user emotion and expected reaction. Furthermore, [14] investigates and innovates the current interactive kiosk to provide immediate responses and reliable information incorporating an intelligent conversational agent (CA).

The programming challenging of the chatbot has been discussed in [15], authors reviewed the problems that occur when designing chatbot using the techniques and concepts of machine learning and natural processing. In [16] the authors proposed a Chatbot for tourism purpose, in which the tourist have limited time. Their methodology divided to three phases. Data preparing was the first phase, which include collecting the famous places and their location in the city. Then mining the collected data and finally use the Dialog Flow. In [17] the authors proposed Chatbot for Medical consultant (MedBot), they implement their Chatbot in IM application using the line application. Figure 2 presents the system architecture of the MedBot. First, the user transfer the message, then the message will be transfer to the application, finally the application will forward the received message to the Dialogflow; is the engine of the their Chatbot.

Although Chatbotis used in many fields, to the best knowledge of the researcher, it is not yet used in the field of information security. In the field of information security, prompt, accurate, and secure responses are highly required. Therefore, this paper is devoted to study the use of, and development of an elegant Chatbot that efficiently and securely provides suitable advice in the field of information security.

## 3. The proposed Model

A. Motivation and Problem Statement

The number of technologies users as well as users of important and sensitive applications, such as banking, is tremendously increasing. Many of those users have limited knowledge about and/or lack most of necessary aspects of information security. Thus, such kind of users are in a dangerous situation since they are accessible for many of bad people. Furthermore, those users are existing in an environment where cybercrimes continuously increase. Moreover, criminal tools and programs are advanced and continuously improved, therefore the need for a protection is continuous and needs some consultation from information security experts.

Provision of consultancy services in a vital field such as information security is an important and plays main role in the raise of awareness about information security for many users. Raising the awareness of users in information alleviate security threats that face them, help them to take the right action to protect themselves, and their society. Having a good level of security in many parts can be achieved through the utilization of chatbot. Therefore, the researchers sought to provide intelligent virtual information security adviser with efficient and accurate replies to the customer.

To best of our knowledge and after a wide read on the previous published researches work, we found many of the chatbots that provide service in many domains such as medical consultant, airline help, tourism and so many other application, but unfortunately we couldn't find any Chatbot that help users to get advices in information security.

B. Proposed solutions

In this paper, we proposed a Chatbot that uses virtual adviser from a json file, which we called "ChatBot Sec" or simply CBS, to offer information security advices through a unified interface. Having such chatbot ("CBS") may increase the reliability and offer accurate advices (collected from json file in a tree structure). The main architecture is depicted on figure3.

Up on a receive of a user's request, that is step 1 in figure 3, the recipient virtual adviser checks to see if there is cached reply for a similar previous request. If so, reply is return to the user. If there was no cached reply, the knowledge base is searched based on the extracted keywords. After that, matching replies are returned to the virtual adviser that conducted search operation. The virtual adviser (which received user's request) formulates (put replies in a certain format) a reply, and sends the reply to the user, and finally caches it (reply) for future similar request. For more explanation, these steps are shown in the below flowchart, figure4.

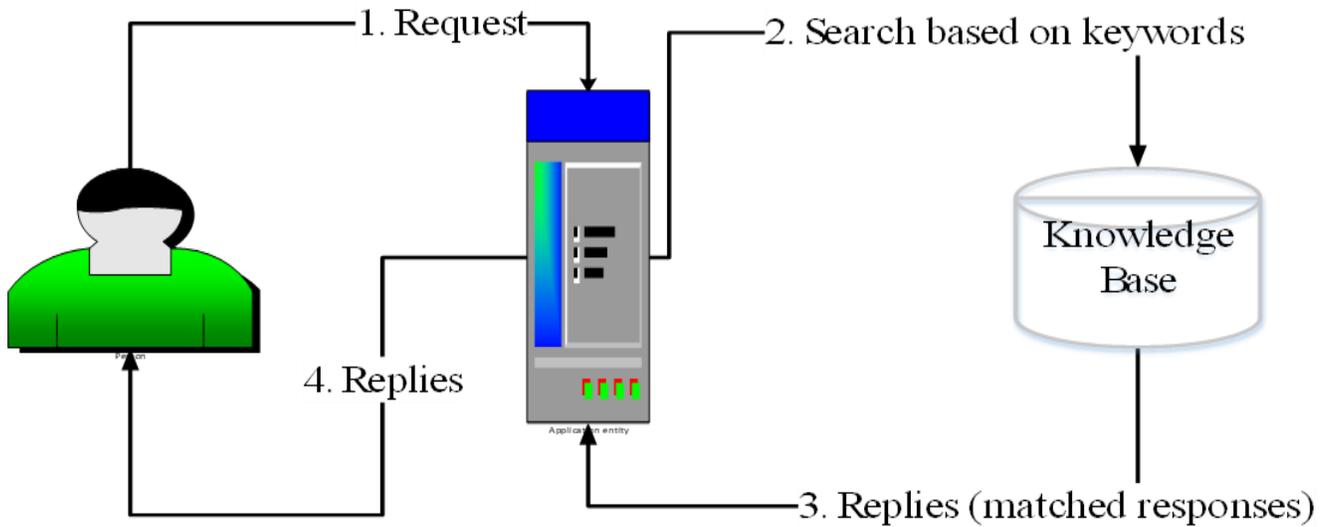

Fig. 3  Architecture of the proposed Chatbot (CBS

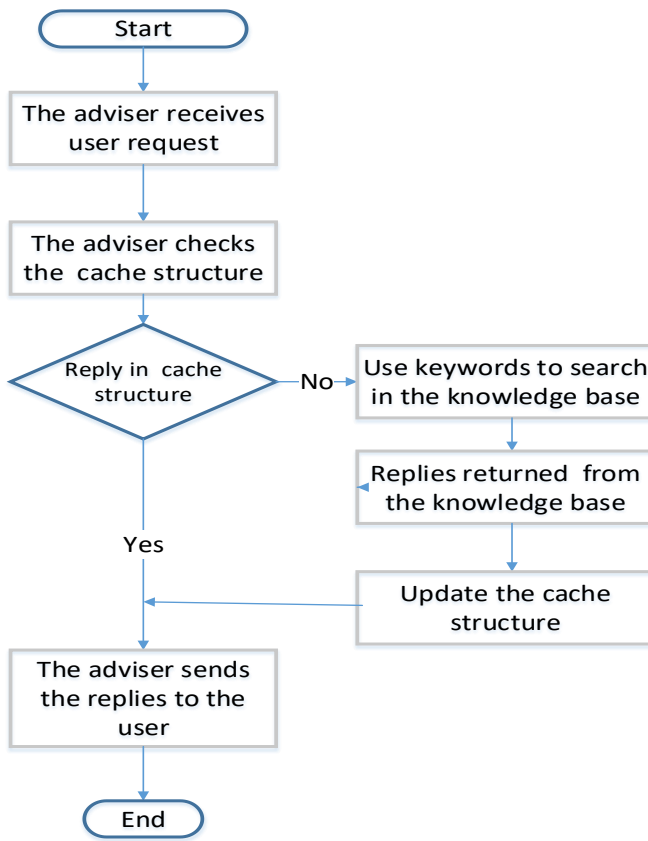

Fig. 4  Actions followed by the proposed Chatbot (CBS)

C. The main advantages of this CBS includes:

- Raising the awareness in a vital field.
- offering advices from a knowledge base that have many different parts, part for each different topic, e.g., part for malware, good and recommended practice, and selection of strong passwords.
- Increase the ability to offer most updated advice
- Fast response for similar queries

## 4. Conclusions and future work

In this paper, we have developed a chatbot that extracts keywords during a chat with users, then uses the keywords to search a knowledge base (in a json file) and sends matching replies for users. Having such chatbot, many advantages have been achieved. These advantages include raising the awareness in a vital field, which reduces security threats normal users face. In addition, provision of an accurate and fast response. This chatbot is intended to be deployed over platform of telegram at the beginning. The future work will focus on deployment on more platforms using Json.

**Acknowledgment**

This project was funded by Deanship of Scientific Research, Northern Border University for their financial support under grant no. SCI-2019-1-10-F-8309. The authors, therefore, acknowledge with thanks DSR technical and financial support.